\documentclass[aps,prl,showpacs,notitlepage,reprint]{revtex4-1}
\usepackage{graphicx}
\usepackage{dcolumn}
\usepackage{amsmath}
\usepackage{bm}
\usepackage{todonotes}
\usepackage[utf8]{inputenc}
\usepackage{amsmath}
\usepackage{natbib}
\usepackage{graphicx}
\usepackage{caption}
\usepackage{subcaption}
\usepackage{lipsum}
\usepackage{float}

\newcommand{\bn}[1]{\mbox{\boldmath$#1$}}

\newcommand{\beq}{\begin{equation}}
\newcommand{\eeq}{\end {equation}}
\newcommand{\bea}{\begin{eqnarray}}
\newcommand{\eea}{\end{eqnarray}}

\begin{document}
\title{Optical properties of radially-polarised twisted light}

\author{K. Koksal$^{1,2}$, {\rm {M. Babiker}}$^{1}$}
\affiliation{$^{1}$Department of Physics, University of York, YO10 5DD, UK} 
\affiliation{$^2$Physics Department, Bitlis Eren University, Bitlis 13000, Turkey}
 \author{V. E. Lembessis$^{3}$}
 \affiliation{$^3$Quantum Technology Group, Department of Physics and Astronomy, College of Science, King Saud University, Riyadh 11451, Saudi Arabia}
\date{\today}

\begin{abstract}

We show that, in general, any type of radially-polarised paraxial twisted optical mode carries only an axial total optical angular momentum (AM) ${\bar {\cal J}_z}=\ell{\cal L}_0$ where $\ell$ is  the winding number and ${\cal L}_0$ is a constant.  This mode, however, is shown to have zero spin angular momentum (SAM), so it is endowed only with orbital angular momentum (OAM) and no SAM. The helicity is found to be proportional to $\ell$, hence radially-polarised modes display chirality. When applied to  a Laguerre-Gaussian (LG) mode our treatment leads to a total helicity equal to $(\ell/|\ell|){\cal Q}$, where ${\cal Q}$ is the action constant. The factor $(\ell/|\ell|)=\pm 1$, depends on the sign, not the magnitude of $\ell$ and so the result holds for any radially-polarised LG mode however large the magnitude of its winding number $\ell$ is.  The magnitude of the action constant ${\cal Q}$ and hence the helicity are diminished for all such LG modes of large beam waist $w_0$.
\end{abstract}


\maketitle

\section{Introduction}
A great deal of work has already been carried out on twisted light and on exploring its interaction with matter  \cite{allen1999,lesallen2003,zhan2004,frank2008,Andrews2012,torres2011}, including atoms and molecules \cite{babiker2018atoms}.  Cylindrical twisted light modes come in a variety of forms, depending on how they are generated, but all the familiar forms are commonly characterised by the ubiquitous azimuthal phase factor $e^{i\ell\phi}$ in the generic amplitude function, which we denote by ${\cal F}$.  Here $\ell$ is the winding number and $\phi$ is the azimuthal angle in cylindrical polar coordinates. Wave polarisation is an additional, equally significant, ingredient adding another layer of complexity to twisted light.  Familiar wave polarisation types include linear, circular and elliptical, together with superpositions of those when dealing with interfering modes.  Other forms of polarisation are generally referred to as vector modes \cite{Holmes2019} and include, in particular, the radial and azimuthal polarisations and superpositions.  Here we are concerned with radially-polarised modes which, since their generation in laser oscillations \cite{mushiake1972,kozawa2005}, have been of considerable interest and, especially so, more recently in the context of twisted light \cite{zhan2009}.  A prominent characteristic of radially-polarised light is that its beams focus into very small waists compared with uniformly-polarised modes \cite{Kozawa2007,dorn2003,bash2010,levy2019}.

However, as far as we know, the general properties of radially-polarised twisted light have not been explored. In particular, there is need to determine the energy-momentum, spin angular momentum, total angular momentum and the helicity and chirality of such modes.  The purpose of this article is to set out the formalism needed to determine these properties.  We focus on the paraxial regime and aim to evaluate the properties to leading order.  The main ingredients of the formalism emphasise the need to ensure that the electromagnetic fields include the longitudinal components for both the electric and magnetic fields and that Maxwell's equations are satisfied.  This means that once the magnetic field format is determined, the electric field follows by application of Maxwell's curl equation and vice versa the magnetic field follows from the electric field by application of the second curl equation. 

We aim to deal with monochromatic paraxial radially-polarised vortex modes, mainly without specifying the kind of vortex mode and consider the  cycle-averaged properties. These are the spin angular momentum, the orbital angular momentum and the helicity and chirality.  We aim to show  how the general results for an arbitrary optical vortex lead to well-defined properties when the type of vortex mode is specified and we evaluate the properties for the special case of a Laguerre-Gaussian optical vortex mode.

We find that, typically any radially-polarised paraxial twisted optical mode is endowed with a total angular momentum (AM) which is only $\ell$ quantised where $\ell$ is  the winding number and it always has zero spin angular momentum (SAM), so it just carries orbital angular momentum (OAM) and no SAM. The optical helicity is shown to be proportional to $\ell$, confirming that radially-polarised modes display chirality as it changes sign with the sign of the winding number $\ell$. This is a manifestation of geometrical chirality arising from the geometrical structure of the beam \cite{nesh2021}. Our general results are then applied to the case of  a Laguerre-Gaussian (LG) mode and we find that the total helicity is equal to $(\ell/|\ell|){\cal Q}$, where $Q$ is the action constant, while the factor $(\ell/|\ell|)=\pm 1$ is the Hopf index of the radially-polarise LG mode. We show that this result is applicable to any radially-polarised LG mode however large the magnitude of its winding number $\ell$ is.  We also find that the helicity is diminished for all such LG modes of large beam waist $w_0$.

\section{Radially-polarised vortex fields}
In cylindrical coordinates the electric and magnetic fields of a radially-polarised paraxial twisted light mode are derivable from a vector potential in the form
\beq
{\bf A}(\rho,\phi,z)={\bn {\hat \rho}}{\cal F}_{\ell, m}(\rho,\phi)e^{ik_zz}
\label{vect1}
\eeq
where carets denote unit vectors, $k_z$ is the axial wavevector with the light travelling along the $+z$ axis and ${\cal F}_{\ell, m}$ is the mode function which includes both the amplitude and phase functions in terms of the plane-polar coordinates $(\rho,\phi)$. The mode is  labelled by the indices $\ell$ and  $m$, with $\ell$ the winding number and $m$ could be an azimuthal number, as in the case of Laguerre-Gaussian (LG) optical vortex modes, or could be redundant, as in the case of optical Bessel modes, but the treatment is not restricted to LG or Bessel modes and is applicable in general for other radially-polarised paraxial optical vortex modes.


Using the standard form of the curl of a vector in cylindrical coordinates we obtain for the magnetic field 
${\bf B}(\rho,\phi,z)={\bn {\nabla}}\times{\bf A}(\rho,\phi,z)$.
With the vector potential as defined for the radially-polarised mode in Eq.(\ref{vect1}) we obtain the paraxial  magnetic field to leading order 
\beq
{\bf B}(\rho,\phi,z)=ik_z{\bn {\hat \phi}}{\cal F} e^{ik_zz}-{\bn {\hat z}}\frac{1}{\rho}\frac{\partial {\cal F}}{\partial \phi}e^{i k_z z}
\label{magfield}
\eeq

The electric field  follows from the magnetic field using Maxwell's equation 
\beq
{\bn {\nabla}}\times {\bf B}=\frac{1}{c^2}\frac{\partial{\bf E}}{\partial t}
\eeq
We obtain,  to the same leading order as for the magnetic field
\beq
{\bf E}(\rho,\phi,z)=ick_z{\bn {\hat \rho}}{\cal F}e^{i k_z z}
-{\bn {\hat z}}c\frac{1}{\rho}\frac{\partial(\rho {\cal F})}{\partial \rho}e^{ik_zz}
\label{electfield}
\eeq
The general paraxial radially-polarised vector modes we are concerned with here have the following  mode function 
\beq
{\cal F}_{\ell,m}(\rho,\phi)={\tilde {\cal F}}_{\ell,m}(\rho)e^{i\ell\phi}e^{i k_z z}
\eeq
where we have separated the phase dependence, so now ${\tilde{\cal F}}_{\ell,m}$ is a radial amplitude function (as it depends on only the radial coordinate $\rho$). Since we are dealing with optical vortex modes the index $\ell$ is the familiar  winding number, while $m$ could be a radial number as in Laguerre-Gaussian modes, or it could be redundant as for Bessel modes. We shall develop the analysis for a general ${\tilde{\cal F}}$, which could be appropriate for any optical vortex.

\section{cycle-averaged optical properties}

The optical vortex mode properties of significance which we shall be concerned with here are the three cycle-averaged properties, namely the optical spin angular momentum (SAM) density ${\bar {\bf s}}$, the angular momentum (AM) density ${\bn {\bar j}}$ and the helicity density  ${\bar \eta}$. These cycle-averaged densities are generally defined as follows. 
\bea
{\bn {\bar s}}&=&\frac{1}{4\omega}\Im\left\{[\epsilon_0{\bf E}^*\times{\bf E}]+\frac{1}{\mu_0}[{\bf B}^*\times{\bf B}]\right\};\nonumber\\
&=&{\bn {\bar s}}_{E}+{\bn {\bar s}}_{B}\;\;\;\;\;\;{\rm {(SAM \;density)}}
\label{sama}
\eea
\beq
{\bn {\bar j}}={\bf r}\times {\bar {\bn {\pi}}};\;\;\;\;({\rm {AM\; density}})\label{jbar}
\eeq
\beq
{\bar {\eta}}({\bf r})=-\frac{\epsilon_0 c}{2\omega}\Im{[{\bf E}^*\cdot{\bf B}]},\;\;\;\;({\rm {Helicity\; density}});
\label{hel}
\eeq
where in the above ${\bar {\bn {\pi}}}=\frac{1}{c^2}{\bar {\bf w}}$ is the linear momentum density with ${\bar {\bf w}}= \frac{1}{2\mu_0}\Re[{\bf E}^*\times{\bf B}]$ the energy density. 
The symbols ${\Re}[...]$ and $\Im[...]$ stand for real and imaginary parts of [...] and the superscript * in ${\bf E}^*$ stands for the complex conjugate of ${\bf E}$. As stated above we deal in turn with the evaluations of above densities specifically in relation to the radially-polarised optical vortex modes. The final tasks involve evaluating the total (integrated) properties, namely total helicity, total SAM and total angular momentum with each evaluated as the space integral of the density variations over the x-y plane.
\subsection{Evaluation of SAM}
We begin with the evaluation of the spin angular momentum density.  We have for the electric field part ${\bn {\bar s}}_E$
\bea
{\bn {\bar s}}_E&=&\frac{\epsilon_0}{4\omega}\Im[{\bf E}^*\times{\bf E}]
\label{sama2}
\eea
Substituting for the electric fields , we have  
\begin{widetext}
\bea
{\bn {\bar s}}_{E}&=&\frac{\epsilon_0}{4\omega}\Im\left(-ick_z{\bn {\hat \rho}}{\cal F}^*e^{-i k_z z}-{\bn {\hat z}}c\left\{\frac{{\cal F^*}}{\rho}+\left[\frac{\partial {\cal F}}{\partial \rho}\right]^*\right\}e^{-ik_zz}\right)\times \left(ick_z{\bn {\hat \rho}}{\cal F}e^{i k_z z}-{\bn {\hat z}}c\left\{\frac{{\cal F}}{\rho}+\frac{\partial {\cal F}}{\partial \rho}\right\}e^{ik_zz}\right)\\
&=&-\frac{ c^2 k \epsilon_0}{2\omega} \left(\frac{|{\tilde{\cal F}}|^2}{\rho}+{\tilde{\cal F}}{\tilde{\cal F}}' \right){\bn {\hat \phi}}\nonumber\\
&=&-\frac{ c^2 k \epsilon_0}{2\omega} \left(\frac{|{\tilde{\cal F}}|^2}{\rho}+{\tilde{\cal F}}{\tilde{\cal F}}' \right)(-\sin\phi{\bn {\hat x}}+\cos\phi{\bn {\hat y}})
\label{samE}
\eea
\end{widetext}
Thus we have found that the electric field part of the SAM density of a radially-polarised optical vortex mode is oriented azimuthally, so has both $x-$ and $y-$ components which vary with $\phi$.   

Consider next the magnetic field contribution ${\bn {\bar s}}_B=\frac{1}{4\omega\mu_0}\Im[{\bf B}^*\times{\bf B}]$.  We have
\begin{widetext}
\bea
{\bn {\bar s}}_{B}&=&\frac{1}{4\mu_0\omega}\Im\left\{\left(-ik_z{\bn {\hat \phi}}{\cal F} e^{-ik_zz}-{\bn {\hat z}}\frac{1}{\rho}\left[\frac{\partial {\cal F}}{\partial \phi}\right]^*e^{-i k_z z}\right)\times \left(ik_z{\bn {\hat \phi}}{\cal F} e^{ik_zz}-{\bn {\hat z}}\frac{1}{\rho}\left[\frac{\partial {\cal F}}{\partial \phi}\right]e^{i k_z z}\right)\right\}\nonumber\\
&=&\Im\left( \frac{ik_z}{4\mu_0\omega\rho}\right)\left[{\cal F}^*\left(\frac{\partial {\cal F}}{\partial \phi}\right)+{\cal F}\left(\frac{\partial {\cal F}}{\partial \phi}\right)^*\right]{\bn {\hat \phi}}\times{\bn {\hat z}}={\bf 0}
\eea
\end{widetext}
The last equality follows once we evaluate the angular derivatives.  Thus the magnetic field contribution to the SAM density is zero and we have only the electric field contribution ${\bn {\bar s}}_E$, given by Eq.(\ref{samE}).

The total (space-integrated) SAM vanishes identically 
\beq
{\bn {\bar S}}=\int_{0}^{2\pi}d\phi\int_0^{\infty} \rho\;d\rho\;[{\bn {\bar s}}_E+{\bn {\bar s}}_B]={\bf 0};
\label{sam3}
\eeq
where each of the $x-$ and $y-$ components of the SAM density yields a zero result by virtue of the angular integration.

\subsection{Evaluation of angular momentum}
Next we evaluate the cycle-averaged angular momentum density of the radially-polarised vortex which is given by
\beq
{\bn {\bar j}}={\bf r}\times {\bar {\bn {\pi}}}=\frac{1}{2c^2\mu_0}\left[{\bf r}\times\left(\Re[{\bf E}^*\times{\bf B}\right)\right]
\label{jdens}
\eeq
Consider the cross product in Eq.(\ref{jdens}) which we evaluate as follows
\begin{widetext}
\bea
{\bf E}^*\times{\bf B}&=&\left(-ick_z{\bn {\hat \rho}}{\cal F}^*e^{-i k_z z}-{\bn {\hat z}}c\left\{\frac{{\cal F^*}}{\rho}+\left[\frac{\partial {\cal F}}{\partial \rho}\right]^*\right\}e^{-ik_zz}\right)\times \left(ik_z{\bn {\hat \phi}}{\cal F} e^{ik_zz}-e^{i k_z z}{\bn {\hat z}}\frac{\partial {\cal F}}{\rho\partial \phi}\right)\nonumber\\
&=&i c k_z \left(\frac{|{\tilde{\cal F}}|^2}{\rho}+{\tilde{\cal F}}{\tilde{\cal F}}' \right){\bn {\hat \rho}}+ \left\{\ell ck_z\frac{|{\tilde {\cal F}}|^2}{\rho}\right\}{\bn {\hat \phi}}+k_z^2 c |{\tilde{\cal F}}|^2{\bn {\hat z}}
\label{sam2}
\eea
\end{widetext}
Therefore on taking the real part, we have
\beq
\Re[{\bf E}^*\times{\bf B}]=k_z c \ell\frac{|{\tilde{\cal F}}|^2}{\rho} {\bn {\hat \phi}}+k_z^2 c |{\tilde {\cal F}}|^2{\bn {\hat z}}
\eeq
The angular momentum density follows
\bea
{\bn {\bar j}}&=&\frac{1}{2 c^2\mu_0}{\bf r}\times \Re[{\bf E}^*\times{\bf B}]\nonumber\\
&=& \left(\frac{k_z}{2 c\mu_0}\right) \{\rho\bn{{\hat \rho}}\}\times\left\{\ell\frac{|{\tilde{\cal F}}|^2}{\rho}{\bn {\hat \phi}}+k_z {|{\tilde{\cal F}}|^2}{\bn {\hat z}}\right\}\nonumber\\
&=& \left(\frac{k_z}{2 c\mu_0}\right) \left\{\ell{|{\tilde{\cal F}}|^2}{\bn {\hat z}}-k_z |{\tilde{\cal F}}|^2\rho{\bn {\hat \phi}}\right\}\nonumber\\
\eea
Since we have ${\bn {\hat \phi}}=-{\bn {\hat x}}\sin\phi+{\bn {\hat y}}\cos\phi$ the angular momentum density vector has all three Cartesian components.  However, the transverse ($x-$ and $y-$) components are $\phi -$ dependent and, as we point out shortly, will result in zero on angular integration.

Once again we consider the total angular momentum as the space integral of the angular momentum  density.  

\beq
{\bn {\bar {\cal J}}}=\int_0^{2\pi}d\phi\int_0^{\infty}\rho\;d\rho\; {\bn {\bar j}}
\eeq
The $x-$ and $y-$ components give zero each due to vanishing angular integration.  We are left only with the z-component, so we have
\beq
{\bn {\bar {\cal J}}}={\bn {\hat z}}\ell\left(\frac{k_z\pi}{c\mu_0}\right)I_P
\label{jbar3}
\eeq
where the integral $I_P$ is related to the applied power $\cal P$ of the mode,  evaluated as the space integral over the beam cross-section of the z-component of the Poynting vector.  We have 
\beq
{\cal P}=\frac{1}{2\mu_0}\int_0^{2\pi}d\phi\int_0^{\infty}|({\bf E}^*\times{\bf B})_z|\rho d\rho
\label{eypee}
\eeq
with
\beq
({\bf E}^*\times{\bf B})_z= c k_z^2 |{\tilde{\cal F}}|^2
\eeq
We can then write for $I_P$
\beq
I_P=\int_0^{\infty}|{\tilde{\cal F}}|^2\rho\;d\rho\label{integ}
\eeq
Thus we obtain for the power ${\cal P}$
\beq
{\cal P}=\left(\frac{\pi ck_z^2}{\mu_0}\right)I_P
\label{calpee}
\eeq
Substituting for $I_P$, we have for the total angular momentum per unit length
\beq
{\bn {\bar {\cal J}}}=\ell{\cal L}_0{\bn {\hat z}}
\eeq
where ${\cal L}_0$ has the dimensions of angular momentum per unit length and is given by
\beq
{\cal L}_0=\left(\frac{{\cal P}}{k_z c^2}\right)\label{ell0}
\eeq
Thus we find that ${\bn {\bar {\cal J}}}$ is axial and proportional to $\ell$ which confirms that the angular momentum carried by the radially-polarised LG mode is purely an orbital angular momentum. Note that we have determined the angular momentum without specifying the type of mode.  The result is therefore general and it agrees with the result for linearly-polarised Laguerre-Gaussian light \cite{Koksalc2022}. There is no contribution to be associated with spin angular momentum, which we have already confirmed to be zero.

\subsection{Evaluation of helicity and chirality}
In optical physics, the properties of helicity and chirality are such that two optical modes which differ only in the sign of the winding number $\ell$ are distinguishable. In that case, one beam is a phase-inverted mirror image of the other beam.  Such kind of optical mode is then said to exhibit chirality which has been a subject of considerable investigation \cite{ranada1989,ranada1995,ranada1996,tang2010,Cameron2012,zamberanapuyalto2012,afanasiev1996}.  More recent investigations have spurred interest in chirality which include the references \cite{bliokh2017,inproceedings,inproceedings2,wozniak2018,nesh2021,lembessis2021,forbes2021}.

In their recent work, Nechayev et al \cite{nesh2021} suggested that there exist two kinds of chirality.  The first is non-geometrical which includes optical chirality due to elliptical polarisation and the second, which is termed the Kelvin chirality, depends on the geometrical structure of the mode. 
This work followed the experiment by Wozniak et al \cite{wozniak2018} in which linearly-polarised Laguerre-Gaussian modes were shown to display chirality.
Here we seek to evaluate the chirality density and its space integral and aim to find out whether the chirality of radially-polarised optical modes can be classed as Kelvin chirality, or is it the usual optical non-geometrical chirality.

We begin by considering helicity, rather than chirality, as the two differ by a proportionality factor \cite{Koksal2022}.  The  cycle-averaged helicity density of the radially-polarised mode is as defined generally in Eq.(\ref{hel}). Substituting for the fields using Eqs.(\ref{magfield}) and (\ref{electfield}), we have for the dot product
\begin{widetext}
\bea  
[{\bf E}^*\cdot{\bf B}]&=&\left\{-ick_z{\bn {\hat \rho}}{\cal F}e^{-i k_z z}
-{\bn {\hat z}}c\frac{1}{\rho}\left(\frac{\partial(\rho {\cal F})}{\partial \rho}\right)^*e^{-ik_zz}\right\}\cdot\left\{ik_z{\bn {\hat \phi}}{\cal F} e^{ik_zz}-{\bn {\hat z}}\frac{1}{\rho}\frac{\partial {\cal F}}{\partial \phi}e^{i k_z z}\right\}\nonumber\\
&=&c\frac{1}{\rho^2}\left(\frac{\partial(\rho {\cal F})}{\partial \rho}\right)^*\left(\frac{\partial {\cal F}}{\partial \phi}\right)\nonumber\\
&=&\frac{ic\ell}{\rho^2}\left(\frac{\partial(\rho {\tilde {\cal F})}}{\partial \rho}\right){\tilde{\cal F}}=i\ell c\left[\frac{1}{\rho}{\tilde{\cal F}}'{\tilde{\cal F}}+\frac{1}{\rho^2}|{\tilde {\cal F}}|^2\right]
\eea
\end{widetext}
where ${\tilde{\cal F}'}=d{\tilde{\cal F}}/d\rho$. We therefore have for the helicity density of the radially-polarised general vortex mode
\beq
{\bar {\eta}}({\bf r})=-\ell\frac{\epsilon_0 c^2}{2\omega}\left[\frac{1}{\rho}{\tilde{\cal F}'}{\tilde{\cal F}}+\frac{1}{\rho^2}|{\tilde {\cal F}}|^2\right]
\label{heldens1}
\eeq
This result is applicable to any paraxial radially-polarised optical vortex and, if required for a particular case, all we need then is to specify the amplitude function ${\tilde {\cal F}}$.  Note that the helicity density is proportional to $\ell$ and so in addition to changing with the magnitude of $\ell$, it also changes with the sign of $\ell$, hence exhibiting the chirality feature that is common to all optical vortex modes. 

\subsection{Integrated Helicity}

The total integral of the helicity density over the $x-y$ plane is 
\bea
{\cal {\bar C}}_{\ell, p}
&=&-\ell \frac{\pi\epsilon_0 c^2}{\omega}\int_{0}^{\infty} \rho d\rho\left[\frac{1}{\rho}{\tilde{\cal F}}_{p\ell}'{\tilde{\cal F}}_{p\ell }+\frac{1}{\rho^2}|{\tilde {\cal F}}_{p\ell}|^2\right]\nonumber\\
&=& {\cal I}_1+{\cal I}_2
\label{integrands}
\eea
We now show that ${\cal I}_1$ is identically zero for all ${\tilde {\cal F}}$.  We have
\bea
{\cal I}_1&=&-\ell\frac{\epsilon_0\pi c^2}{2\omega}\int_{0}^{\infty} \left(\frac{d}{d\rho}{\tilde {\cal F}}^2\right)\;d\rho\nonumber\\
&=&-\ell\frac{\epsilon_0\pi c^2}{2\omega}\left[{\tilde {\cal F}}^2(\rho)\right]_0^{\infty}\nonumber\\
&=&0\label{zero}
\eea
since the ${\tilde {\cal F}}(0)=0={\tilde {\cal F}}(\infty)$.
We are thus left with the second term so that for any ${\tilde {\cal F}}$ the helicity per unit length is given by
\beq
{\cal {\bar C}}_{\ell,p}={\cal I}_2
=-\ell\frac{\pi\epsilon_0 c^2}{\omega}\int_{0}^{\infty}\rho d\rho\left[\frac{1}{\rho^2}|{\tilde {\cal F}}_{\ell,p}|^2\right]
\label{ceebar1}
\eeq
This is the general expression for the total helicity per unit length of a mode of any paraxial twisted light mode that is radially-polarised.  Although we are able to make definitive statements about the total SAM and total angular momentum for all forms of ${\cal F}$ characterising radially-polarised twisted light, we are not able to proceed further to evaluate the  total helicity in Eq.(\ref{ceebar1}) without specific knowledge about the form of ${\tilde {\cal F}}$.  In the next section we focus on the form of ${\tilde {\cal F}}$ appropriate for a Laguerre-Gaussian optical vortex and aim to evaluate the helicity density given by Eq.(\ref{heldens1}) and the total helicity given by  Eq.(\ref{ceebar1}).  

\section{Applications to Laguerre-Gaussian modes}

 A paraxial Laguerre-Gaussian mode of winding number $\ell$, radial number $p$ and waist $w_0$ has an amplitude function given by
\beq
{\tilde {\cal F}}_{\ell,p}(\rho)={\cal E}_0\sqrt{\frac{p!}{(p+|\ell|)!}} e^{-\frac{\rho^2}{w_0^2}}  \left(\frac{\sqrt{2}\rho}{w_0}\right)^{|\ell| }L^{|\ell|}_p\left(\frac{2\rho^2}{w_0^2}\right)\label{efftilde}
\eeq
where ${\cal E}_0$ is a normalisation factor and we have identified $m$ as the radial number $p$ in LG modes.  The factor ${\cal E}_0$ is determined in terms of the applied power $\cal P$ of the mode which we have already evaluated above in terms of the integral $I_P$.  For a Laguerre-Gaussian mode the integral in (\ref{integ}) is standard and gives
\beq 
I_P=\int_0^{\infty}|{\tilde{\cal F}}|^2\rho\;d\rho=\frac{1}{4}{\cal E}_0^2w_0^2
\eeq
and so we can now determine the overall factor ${\cal E}_0$ for the Laguerre-Gaussian beam.  We have 
\beq 
{\cal E}_0^2=\frac{4{\cal P}}{\pi\epsilon_0 c^3k_z^2w_0^2}\label{ee0}
\eeq
Consider first the variations of the helicity density for representative modes, namely $\ell=1,2$.  The general expression for the helicity density Eq.(\ref{heldens1}) is circularly-symmetric as it is a function only of the radial coordinate $\rho$.   For a LG mode, we simply substitute for ${\tilde {\cal F}}$ given by Eq.(\ref{efftilde}) and we obtain the variation of the helicity density, as shown in Fig. 1 for $\ell=1$ and Fig. 2 for $\ell=2$. The plots in each case show the contributions from the first term, the second term and their sum. Confirmation of the vanishing integral of the first term are shown in  Figs. 3 for $\ell=1$ and Fig.4 for $\ell=2$ where the areas under the curves corresponding to the areas enclosed by the  integrands due to the first term are zero for both $\ell=1$ and $\ell=2$. We have confirmed that the variations shown for the case $\ell=2$ define the trend for $\ell> 2$

Note, in particular, that the variations of the helicity density for the case $\ell=1$, shown in Fig. 1 differ significantly from those of $\ell\geq 2$ in Fig.2, primarily in that the helicity density does not vanish at the core where $\rho=0$ for $\ell=1$, while it does vanish at $\rho=0$ for $\ell\geq 2$.  This behaviour can be explained by inspecting the general form of the helicity density Eq.(\ref{heldens1}).  When applied to the Laguerre-Gaussian ${\cal F}$ for $\ell=1$, we have from Eq.(\ref{efftilde})
\beq
{\tilde {\cal F}}_{\ell=1}\propto \rho e^{-\rho^2/w_0^2}L^{1}_p\left(\frac{2\rho^2}{w_0^2}\right)  \label{efftildeone}
\eeq
Hence 
\beq 
|{\tilde {\cal F}}_{\ell=1}|^2\propto \rho^2 e^{-2\rho^2/w_0^2}\left[L^{1}_p\left(\frac{2\rho^2}{w_0^2}\right)\right]^2
\eeq
Also, since ${\cal F}'$ does not vanish at $\rho=0$ we can write
\beq
[{\tilde {\cal F}}'{\tilde {\cal F}}]_{\ell=1}\propto \rho e^{-\rho^2/w_0^2}L^{1}_p\left(\frac{2\rho^2}{w_0^2}\right) {\tilde {\cal F}}'_{\ell=1}
\eeq 
When substituted in the helicity density expression Eq.(\ref{heldens1}) we see that the $1/\rho$ in the first term cancels with the factor $\rho$ in the numerator.  Similarly the factor $1/\rho^2$ in the second term cancels the factor $\rho^2$ in the numerator of the second term.  The overall variation amounts to a non zero value of the helicity at $\rho=0$ only in the case $\ell=1$.  This variation contrasts with the case $\ell\geq 2$ in which the numerators in the two terms have higher powers of $\rho$, always guaranteeing that  the helicity density vanishes at $\rho=0$. 

\begin{figure}
\includegraphics[width=.8\linewidth]{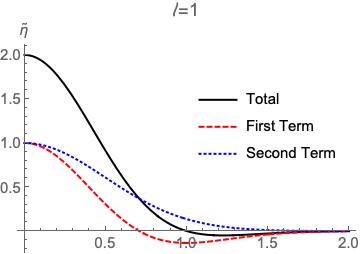}
\caption{Variation with the radial coordinate $\rho$ (in units of the beam waist $w_0$) of the helicity density due to LG modes for which (a) $\ell=1$.  The dashed red curve shows the contribution of the first term in Eq.(\ref{heldens1}) involving the derivative ${\tilde {\cal F}'}$. The blue dotted curve represents the contribution of the second term and the solid black curve is the sum. Note in particular that for $\ell=1$ the helicity density does not vanish at the core $\rho=0$}.
\end{figure}

\begin{figure}
\includegraphics[width=.8\linewidth]{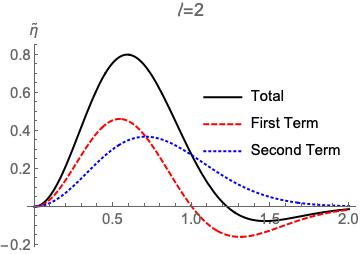}
\caption{The case of LG mode with $\ell=2$. Variation with the radial coordinate $\rho$ (in units of the beam waist $w_0$) of the helicity density due to LG modes for which (a) $\ell=2$.  The dashed red curve shows the contribution of the first term in Eq.(\ref{heldens1}) involving the derivative ${\tilde {\cal F}'}$. The blue dotted curve represents the contribution of the second term and the solid black curve is the sum. Note in particular that for $\ell=2$ the helicity density vanishes at the core $\rho=0$}.
\end{figure}

\begin{figure}
\includegraphics[width=.8\linewidth]{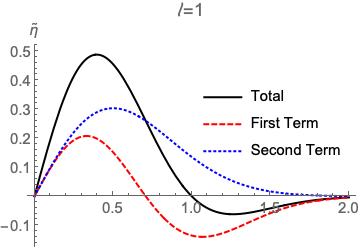}
\caption{The case of LG mode with $\ell=1$. Variations of the helicity density integrand terms in ${\cal I}_1$ and ${\cal I}_2$, (defined in Eq.(\ref{integrands})) with the radial coordinate (in units of $w_0$).  The area enclosed by each curve corresponds to the contribution of the term to the total (integrated) helicity.  The area enclosed by the dashed red curve is verified to be zero, consistent with Eq.(\ref{zero}).}
\end{figure}

\begin{figure}
\includegraphics[width=.8\linewidth]{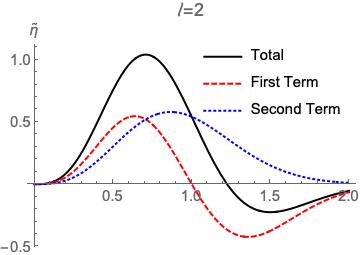}
\caption{The case of LG mode with $\ell=2$. Variations of the helicity density integrand terms in ${\cal I}_1$ and ${\cal I}_2$, (defined in Eq.(\ref{integrands})) with the radial coordinate (in units of $w_0$).  The area enclosed by each curve corresponds to the contribution of the term to the total (integrated) helicity.  The area enclosed by the dashed red curve is verified to be zero, consistent with Eq.(\ref{zero}).}
\end{figure}

It is straightforward to proceed to evaluate the helicity per unit length for ${\tilde {\cal F}}$ corresponding to a Laguerre-Gaussian mode. Using the integration variable $x=2\rho^2/w_0^2$ we have
\bea
{\cal {\bar C}}_{\ell, p}&=&-\ell\frac{\pi\epsilon_0 c^2}{\omega}{\cal E}_0^2\frac{p!}{2(p+|\ell|)!}\int_0^{\infty} x^{|\ell|-1}e^{-x}[L^{|\ell|}_p(x)]^2dx\nonumber\\
&=&-\ell\frac{\pi\epsilon_0 c^2}{\omega}{\cal E}_0^2\frac{1}{2|\ell|}\label{ceebar}
\eea
The details of the evaluation of the integral in Eq.(\ref{ceebar}) are shown in Appendix A. We now have
\bea 
{\cal {\bar C}}_{\ell,p}&=&-\frac{{\cal E}_0^2\pi  \epsilon_0 c^2}{2 \omega}\frac{\ell}{|\ell|}\nonumber\\
&=&-\left(\frac{\ell}{|\ell|}\right)\left(\frac{{\cal P}}{k_z c^2}\right)\left\{\frac{2}{k_z^2w_0^2}\right\}\nonumber\\
&=&\pm{\cal L}_0\left\{\frac{2}{k_z^2w_0^2}\right\}
\eea
where ${\cal L}_0$ is the constant angular momentum per unit length as defined in Eq.(\ref{ell0}) for a fixed power ${\cal P}$ and we have substituted for ${\cal E}_0$ using Eq.(\ref{ee0}). Note that this result is independent of $p$ and clearly depends only on the sign (not the magnitude) of $\ell$. We know that the mode is not circularly polarised, but we have found that the total helicity is similar to, but not the same as, that of circular polarisation, characterised by the pre-factor $\sigma=\pm 1$.  It is easy to check that the helicity has the dimensions of angular momentum per unit length, but the factor $1/k_z^2w_0^2$ is small for $w_0^2>>1/k_z^2$ (which amounts to $w_0>>{\bar {\lambda}}$ where ${\bar {\lambda}}=\lambda/2\pi$ is a reduced wavelength). Thus the helicity is significant in the case of LG beams only for small beam waits $w_0$ and diminishes for progressively larger $w_0$.

\section{Comments and Conclusions}
Our primary aim in this paper involved the derivation of the optical properties of paraxial radially-polarised twisted light modes, namely the spin orbital angular momentum (SAM), the total angular momentum (which is for mally the sum of spin and orbital angular momentum for paraxial light), and their helicity and chirality.  We set out to keep the type of mode unspecified and arrived at the results for the angular momentum and SAM, but for the helicity we are able to arrive at general results and applied them to the Laguerre-Gaussian modes as a specific case. Our treatment is based on general expressions describing the radially-polarised electric and magnetic fields which incorporate the longitudinal component and which were subject to verification of Maxwell's consistency conditions, namely that the electric field in Cartesian coordinates follows from a derived expression of the magnetic field using the Maxwell curl equation, also in Cartesian coordinates, and the magnetic field follows from the electric field using the other Maxwell curl equation. The final Cartesian expressions are then presented in cylindrical polar coordinates. We have checked that the formalism presented is verifiable to leading order in the paraxial approximation. 

We have found that in general, the radially-polarised twisted light modes exhibit only   cycle-averaged transverse SAM components, which arise entirely from the electric field part, are $\phi$-dependent, while the magnetic field  contribution to the SAM density is shown to be identically zero.  The space integral of the SAM density leading to the total SAM is therefore zero. Thus we have confirmed that in general such radially-polarised modes have no SAM. Next we evaluated the angular momentum, which, for paraxial modes, is always the sum of the spin angular momentum and the orbital angular momentum and we have found that the angular momentum density is proportional to the winding number $\ell$, indicating that this contribution of the total angular momentum density is purely orbital, but there are also transverse density components which, like the SAM density case, depend on $\phi$ and so lead to zero on spatial integration. The integrated total angular momentum is also proportional to $\ell$ and so purely orbital in origin. This result is consistent with the SAM result which was evaluated independently, that the mode has no SAM.

The helicity density evaluations could also be carried out for a general radially-polarised optical vortex mode characterised by an unspecified ${\tilde {\cal F}}$ with the density and the general results for the helicity density and the total (integrated) helicity we arrived at are shown in Eq.(\ref{heldens1}) for the helicity density and in Eq.(\ref{ceebar1}) for the integrated helicity.  We proceeded to explore the helicity density variations for the spacial cases of Laguerre-Gaussian modes and pointed out the special behaviour for $\ell=1$ in that the helicity density does not vanish at $\ell=1$ at the core $\rho=0$, while it does vanish for all $\ell\geq 2$.  We explained this behaviour by inspecting the $\rho$ variations of the two terms in the helicity density, confirming that for $\ell=1$ the helicity density has overall dependence $\rho^0$ and so the helicity density does not vanish at $\rho=0$ for $\ell=1$.

Finally, we evaluated the total helicity of the radially-polarised Laguerre-Gaussian optical vortex and found it equal to $(\ell/|\ell|){\cal Q}={\cal N}{\cal Q}$, where ${\cal N}=\pm 1$ is interpreted as a Hopf index and $Q$ is the action constant.  The $\pm 1$ is reminiscent of $\sigma=\pm 1$ for circular polarisation. This result holds for any radially-polarised LG mode however large the magnitude of its winding number $\ell$ is, but the action constant ${\cal Q}$ and consequently the helicity are significant only for small beam waist $w_0$ and  diminish for all such LG modes of large $w_0$. Since this type of helicity (and chirality) originate from the spatial structure of the radially-polarised mode, they can be categorised as of the Kelvin type \cite{nesh2021}.

\subsection{Disclosures}

The authors declare no conflicts of interest.


\begin{widetext}
\section*{Appendix A}

We evaluate the integral in Eq.(\ref{ceebar}) which is
\beq 
{\cal I}=\int_0^{\infty} x^{|\ell|-1}e^{-x}[L^{|\ell|}_p(x)]^2dx
\eeq
Consider the evaluation of following derivative of the function $x^{|\ell|}e^{-x}[L^{|\ell|}_p(x)]^2$ with respect to $x$ 

\beq 
\frac{d}{dx}\{x^{|\ell|}e^{-x}[L^{|\ell|}_p(x)]^2\}=|\ell|x^{|\ell|-1}e^{-x}[L^{|\ell|}_p(x)]^2-x^{|\ell|}e^{-x}[L^{|\ell|}_p(x)]^2-2x^{|\ell|}e^{-x}L^{|\ell|}_p(x)L^{|\ell|+1}_{p-1}(x)
\eeq
So we may now write
\bea 
{\cal I}&=&\int_0^{\infty}x^{|\ell|-1}e^{-x}[L^{|\ell|}_p(x)]^2dx\nonumber\\
&=&\frac{1}{|\ell|}\int_0^{\infty }\left(\frac{d}{dx}\{x^{|\ell|}e^{-x}[L^{|\ell|}_p(x)]^2\}+x^{|\ell|}e^{-x}[L^{|\ell|}_p(x)]^2+2x^{|\ell|}e^{-x}L^{|\ell|}_p(x)L^{|\ell|+1}_{p-1}(x)\right)dx\nonumber\\
&=&\frac{1}{|\ell|}\left([x^{|\ell|}e^{-x}(L^{|\ell|}_p(x))^2]_0^{\infty}+\int_0^{\infty}x^{|\ell|}e^{-x}[L^{|\ell|}_p(x)]^2 dx+2\int_0^{\infty}x^{|\ell|}e^{-x}L^{|\ell|}_p(x)L^{|\ell+1|}_{p-1}(x)dx\right)
\label{integral}
\eea
\end{widetext}
where we have integrated by parts in the derivative term.  The first term in the last equality is zero on applying the integration limits, while the third term is the standard orthogonality integral of the associated Laguerre functions and so vanishes as well.  We are thus left left with 
\beq 
{\cal I}=\frac{1}{|\ell|}\int_0^{\infty}x^{|\ell|}e^{-x}[L^{|\ell|}_p(x)]^2dx=\frac{1}{|\ell|}\frac{(p+|\ell|)!}{p!}\label{result}
\eeq

\bibliography{bibliography.bib}
\end{document}